# Component Importance and Interdependence Analysis for Transmission, Distribution and Communication Systems

Tao Fu, *Member, IEEE*, Dexin Wang, *Member, IEEE*, Xiaoyuan Fan, *Senior Member, IEEE,* and Qiuhua Huang, *Member, IEEE*

*Abstract*—For critical infrastructure restoration planning, the real-time scheduling and coordination of system restoration efforts, the key in decision-making is to prioritize those critical components that are out of service during the restoration. For this purpose, there is a need for component importance analysis. While it has been investigated extensively for individual systems, component importance considering interdependence among transmission, distribution and communication (T&D&C) systems has not been systematically analyzed and widely adopted. In this study, we propose a component importance assessment method in the context of interdependence between T&D&C networks. Analytic methods for multilayer networks and a set of metrics have been applied for assessing the component importance and interdependence between T&D&C networks based on their physical characteristics. The proposed methodology is further validated with integrated synthetic Illinois regional transmission, distribution, and communication (T&D&C) systems, the results reveal the unique characteristics of component/node importance, which may be strongly affected by the network topology and cross-domain node mapping.

## NOMENCLATURE

| | |
|---|---|
| $T$ | Transmission network |
| $D$ | Distribution network |
| $C$ | Communication network |
| DER | Distributed energy resource |
| CNA | Complex network analysis |
| $G$ | Graph of a network |
| $N$ | Number of nodes in a graph |
| $L$ | Number of edges in a graph |
| $V$ | Node/vertex set in a graph |
| $v$ | A single node/vertex in a graph |
| $E$ | Edges in a graph |
| $E_{ij}$ | Edge between nodes $i$ and $j$ |
| $A$ | Adjacency matrix of a graph |
| $i, j$ | Node/vertex index |
| $a_{ij}$ | Element in an adjacency matrix for nodes $i$ and $j$ |
| $Z_{ij}$ | Impedance between nodes $i$ and $j$ |
| $R_{ij}$ | Resistance between nodes $i$ and $j$ |
| $X_{ij}$ | Reactance between nodes $i$ and $j$ |
| $I_{ij}$ | Magnitude of impedance between nodes $i$ and $j$ |
| $w_{ij}$ | Normalized edge weight between nodes $i$ and $j$ |
| $c_v^j$ | Cross-closeness centrality |
| $d_{pq}$ | Shortest path between vertices $p$ and $q$ |
| $G_i, G_j$ | Graph of two different networks |
| $N_i$ | Number of nodes/vertices in graph $G_i$ |
| $N_j$ | Number of nodes/vertices in graph $G_j$ |
| $V_i$ | Node/vertex set in graph $G_i$ |
| $V_j$ | Node/vertex set in graph $G_j$ |
| $p$ | Node/vertex in graph $G_i$ |
| $q$ | Node/vertex in graph $G_j$ |
| $b_v^j$ | Cross-betweenness centrality |
| $\sigma_{pq}$ | Total number of shortest paths between $p$ and $q$ |
| $\sigma_{pq}(v)$ | Number of shortest paths between $p$ and $q$ that include $v$ |
| $E_G$ | Cross-efficiency between graph $G_i$ and graph $G_j$ |
| $V_p$ | Voltage of node $p$ |
| $k_{pq}$ | Coefficient between nodes $p$ and $q$ for cross-efficiency calculation |
| $\Delta E_v(v)$ | Node efficiency drop |
| $E_G(V - v)$ | Cross-efficiency after the removal of the vertex |
| $E(G)$ | Network efficiency of graph $G$ |
| $<k>$ | Average degree of a node in a graph |
| $k_{max}$ | Max degree of a node in a graph |

## I. INTRODUCTION

IDENTIFYING critical components in a system could help plan or prioritize the restoration procedure for distributed infrastructure systems, including transportation networks [1] and power systems [2], as well as other applications like structural vulnerability assessment and cascading blackouts prediction [3]. However, previous efforts and analyses were limited to one stand-alone system. There is a lack of systematic analysis and understanding of component importance and interdependency for power transmission and distribution systems, and their interdependent infrastructures, particularly in the context of system restoration. This motivates us to perform this research to support coordinated *T&D&C* system restoration decision-making.

A significant portion of critical component assessments have been focused on analyzing the component structural importance using network science or complex network theory [2]. In these efforts, infrastructure networks are modeled as graphs or networks, with components being modeled as either nodes or edges. Important components are identified based on their topological features [1]. For the power grid systems, complex network analysis has revealed that most of them are scale-free networks, in which the majority of nodes only have a few links and only few nodes have higher degrees. The most widely used metrics include the degree centrality and betweenness centrality [4]. Following the blackout event in 2003, Albert et al. [5] studied the North American power grid from a network perspective. Using the connectivity loss as the measurement, they showed that failure of only 4% of the high degree or high load nodes could result in up to 60% of connectivity loss, and the loss could be even higher with the consideration of cascading failures. Kinney et al. [6] used network global efficiency between generators and distribution substations as the metric. They demonstrated that the loss of one critical substation with high betweenness and high degree in the North American power grid could lead to 25% loss of transmission efficiency with the triggering of the cascading effects. Crucitti et al. [7] came to the similar conclusion studying the Italian electric power grid. They found that the system can be very vulnerable to the failures that occur on the nodes with highest betweenness centrality. It should be noted that these works focus on the transmission system only. With increasing penetration of distributed energy resources (DERs) and blurred boundary between *T&D* systems, it is imperative to include distribution systems and DERs in critical component analysis.

One of the main shortcomings of pure complex network analysis (CNA) approach is that only topology or connection information is considered, while the physical or functional characteristics of components and systems are ignored. Recent studies have been focusing on extended approaches that incorporate physical properties of power grids into complex network analysis, such as impedance, max power, power flow directions, and simulating a power grid as a directed and weighted graph ([8], [9]). For example, Wei and Liu [10] proposed the combination of impedance and line voltage level for line weights. They showed that their model was more effective in identifying the vulnerable part of a network.

Many of pure CNA and extended CNA studies are mainly focused on a single, non-interacting infrastructure system. Yet, modern systems and critical infrastructures are highly interconnected and mutually dependent in complex ways [11 - 14]. The power incidents also disrupted several key industries, and stressed the entire U.S. Western Interconnection, causing far-reaching security and reliability concerns [11]. On the other hand, electric power infrastructure also depends significantly on other critical infrastructures. Since early 2000, researchers from different domains and areas have devoted significant efforts to identifying, understanding and leveraging interdependency between different networks and infrastructures. Donges et al. [15] investigated the topology of interacting networks, especially the importance and role of single vertex for the interaction or communication between different subnetworks, in a systematic and comprehensive way. A graph-theoretical framework for investigating the topology of interacting networks was proposed, and selected local and global network measures by extending the classical centrality measures such as degree, closeness, betweenness, and cluster coefficient were defined to quantify and investigate the interaction topology of networks on different topological scales. The proposed metrics were used to quantify the structural role of single vertices with respect to the interactions between a pair of subnetworks on both local and global scales. From a risk and vulnerability analysis perspective, Wang et al. [16] pointed out that features of the infrastructure systems must be considered from the perspective of global analysis. They proposed a method for ranking critical components in interdependent infrastructures for protection purposes. In their approach, the importance of an element was evaluated by considering the performance degradation of the network due to the disconnection of the component. The network performance was measured by global safety efficiency of the network $S(G)$, component outage results in a new system denoted by $S(G^*)$. The importance of a component can be assessed by the global efficiency drop after the removing of this component: $\Delta S(C) = (S(G) - S(G^*))/S(G)$. Using the similar approach, Milanovic and Zhu [17] defined the cross-efficiency drop for the coupled electric power system (EPS) network and information and communication technology (ICT) network. They proposed a three-dimensional model, which uses unidirectional edges for power flows and bidirectional edges for information exchanges and demonstrated that efficiency related metrics were most effective in identifying critical interconnected components. However, they only focused on *T&C* systems, and did not consider the physical properties and importance differences of EPS nodes of different nominal voltage levels. Furthermore, they calculated the metrics for each individual network and summed them up as a global metric, which means they did not explicitly consider cross-network interdependence.

This study aims to fill the gap by leveraging the recent development in the field of multilayer networks and several emerging analysis methods, and extend our previous work studying the interdependency between the transmission and communication systems (*T&C*) [18]. Our contributions are:

1) We for the first time included *D* network in interdependency analysis and proposed a multilayer network representation for *T&D&C* system and utilized it for component importance and interdependent analysis.

2) We developed a new approach for the edge weight calculation to reflect individual network physical properties, incorporating both line impedances and connected node voltages for both the power transmission and distribution networks to better distinguish them. In addition, we extended several single-network-oriented metrics [15] and [17], to multilayer networks to assess the node importance for interdependent systems.

3) We compared these network theoretic analysis results with physical-model-based simulation results, including communication network performance analysis using ns-3 simulation [19] and maximum load delivery analysis using *PowerModelsRestoration.jl* [20, 21] to help understand the effectiveness and limitations of network theoretic analysis and how they could be used in conjunction with physical-based approaches.

The rest of the paper is organized as follows: Section II introduces the multilayer network representation of *T&D&C* networks and multilayer network metrics; Section III describes the details of the *T&D&C* system; Section IV presents the network properties for each network from the complex network perspective; Section V compares the results between complex network analysis, the *ns*-3 simulation, and maximum loading capability analysis; the interdependency between network are elaborated in Section VI; Section VII presents the comparison between the proposed edge weight approach and the traditional approach using *T-only* network; and the conclusions are presented in Section VII.

## II. MULTILAYER NETWORKS AND METRICS FOR ANALYSIS

The emerging and unique features of *T&D&C* systems can be better described and highlighted by directly comparing with previously studied *T-only* and *T&C* systems, as follows:

- Distribution systems are not modeled as passive loads connected to transmission buses but detailed subsystems. Their components have their unique functional and physical properties such as radial structure, low voltage and power, sparingly connected with communication systems. Correspondingly, edge and node weights should be properly designed to distinguish components in *T&D* systems.

- Communication was not considered at the edge of the grid in previous studies. We can capture it with *T&D&C* modeling. Note that compared with *T-network*, there is yet much communication coverage in distribution systems except a few nodes connected with *C-network*. These nodes essentially become communication hubs in distribution systems, and such hubs are usually not observed in *T-network*.

We have fully considered these features in network representations (mainly edge and node weights), and proposed several important metrics for analyzing the component importance and interdependence in this paper.

Researches in CNA and extended CNA have been focusing on graph topology related centralities: degree, betweenness, closeness, and efficiency [3]. Researches from Kinney et al. [6] showed that the power grid system can be very vulnerable to the failures that occur on the nodes with the highest betweenness centrality or the highest closeness centrality. Jovica et al. [17] showed efficiency metrics, which use system efficiency and system efficiency drop after removing individual, is the one of the most effective in identifying critical interconnected components. We extended basic centralities and used the following metrics in multilayer interdependent networks for the study of interdependency of T&D&C networks:

- *Cross-closeness centrality*: it measures the closeness of the node p in $G_i$ to $G_j$ along shortest paths.

- *Cross-betweenness centrality*: it measures the importance of node v in $G_i$ that is responsible for passing the information from $G_i$ to $G_j$.

- *Node efficiency drop*: it measures the cross-efficiency drop passing information from $G_i$ to $G_j$ after removing the node v in $G_i$.

### A. Multilayer Network Representation of T&D&C Networks

Using complex network theory, *T*, *D*, or *C* networks can be represented using a graph $G = (V, E)$, where $V = \{V_1, V_2, ..., V_N\}$ is the set of $N$ nodes/vertices in $G$ and $E = \{E_1, E_2, ..., E_{NxN}\}$ is the set of edges/links. Adjacency matrix $A = (a_{ij})_{NxN}$ is used to represent the connectivity between nodes $i$ and $j$: $a_{ij} = 1$ (or weight for a weighted graph) if there is an edge between node $i$ and node $j$ and $a_{ij} = 0$ otherwise.

To simulate the interdependency between *T*, *D*, and *C* networks, we also need to represent the inter-layers edges/links between networks. Since a node $i$ in layer α can be connected to any node $j$ in any layer $\beta$ in a multilayer framework, we adopted the following adjacency matrix to add layer information: $a(i, \alpha, j, \beta) = 1$ (or weight for a weighted graph) if node $i$ in layer α is connected to node $j$ in layer β, otherwise 0.

### B. Edge weights

Edge weights are assigned to take into account of physical properties of *T*, *D*, and *C* networks. For intralayer edges that are within the *T* and *D* networks, their weights are calculated using the line impedance: let $Z_{ij} = R_{ij} + jX_{ij}$ be the impedance of the edge between nodes $i$ and $j$, then $I_{ij} = \sqrt{R_{ij}^2 + X_{ij}^2}$ ; the weight for edge $E_{ij}$ is the normalized $I_{ij}$, i.e., $w_{ij} = I_{ij} / mean(I_{ij})$. For edges in the *C* network, their edges are all set to 1. For all interlayer edges between two networks, their weights are set to 0 because the two nodes for interlayer connections are usually co-located (or at least very close) and thus their distance is zero.

### C. Cross-closeness centrality

Cross-closeness centrality measures the topological closeness of a vertex in one network to the other network along the shortest path. It quantifies the efficiency of interaction between the vertex and the other network. A vertex with high cross-closeness is likely to be important for the fast exchange of energy and/or information with a network [15]. For a vertex $p \in V_i$ in $G_i$ the cross-closeness centrality $c_v^j$ between $p$ and $G_j$ is:

$$c_v^j = \frac{N_j}{\sum_{q \in V_j} d_{pq}} \qquad (1)$$

where $d_{vq}$ is the shortest path between vertices $p$ and $q \in G_j$. If no path exists between $p$ and $q$, $d_{pq} = N_i + N_j - 1$ is set as the upper bound, where $N_i$ and $N_j$ are the number of vertices in $G_i$ and $G_j$, respectively. For generality, $d_{pq}$ is not restricted to any order of vertices and may contain any vertices $v \in V_i \cup V_j$ in any order depending on the topology of the network.

### D. Cross-betweenness centrality

For a vertex $v \in V_i$ in $G_i$, cross-betweenness centrality $b_v^j$ indicates the role of $v$ for mediating or controlling interactions/communication between two networks $G_i$ and $G_j$. It is defined as

$$b_v^j = \sum_{p \in V_i, q \in V_j; p,q \neq v} \frac{\sigma_{pq}(v)}{\sigma_{pq}} \quad (2)$$

, where $\sigma_{pq}$ is the total number of shortest paths between $p$ and $q$ and $\sigma_{pq}(v)$ is the number of shortest paths between $p$ and $q$ that include $v$.

Cross-betweenness centrality describes the importance of vertex $v \in V_i$ as a relational hub between two networks $G_i$ and $G_j$. A vertex with high cross-betweenness may serves as a robust transmitter between two networks.

### E. Cross-efficiency and node efficiency drop

Considering two independent networks $G_i$ and $G_j$, cross-efficiency measures how efficiently the information is exchanged between nodes in these two networks. Assuming the information is transmitted along the shortest paths, information exchange between two nodes in two different networks is disproportional to the length of the shortest path between two nodes, i.e., more efficient if the shortest paths are shorter. We adopted the single-network efficiency metric from Milanović and Zhu [17], and further proposed the cross-efficiency between two networks by adding a coefficient $k_{pq}$ to represent the weight for the starting node. The cross-efficiency of the interconnected network $G = (G_i \cup G_j)$ denoted by $E_G$ can be calculated as follows:

$$E_G = \frac{1}{N_i N_j (\sum_{p,q} k_{pq})/N_e} \sum_{p,q} \frac{k_{pq}}{d_{pq}}, \quad k_{pq} = V_p \quad (3)$$

where $d_{pq}$ is the shortest path between vertices $p \in V_i$ and $q \in V_j$, and $N_i$ and $N_j$ are the number of vertices in $G_i$ and $G_j$, respectively. $k_{pq}$ is set to the nominal voltage (e.g., 500 kV) of node $p$: $V_p$, which gives higher weights to nodes with higher voltages. $N_e$ is the number of shortest paths from $p \in V_i$ to $q \in V_j$.

For a vertex $v \in V_i$, the cross-efficiency drop $\Delta E_v(v)$ for $v$ is defined as the drop of the cross-efficiency after removing $v$ from the $V_i$,

$$\Delta E_G(v) = \frac{E_G - E_{G(V-v)}}{E_G} \quad (4)$$

where $E_G(V-v)$ is the cross-efficiency of $G$ after the removal of the vertex $v$.

Cross-efficiency drop for the vertex $v \in V_i$ describes the importance of the $v$ in the interaction between two networks $G_i$ and $G_j$, with higher cross-efficiency drop indicating higher node importance. For example, a vertex with $\Delta E_G(v) = 1$ indicates that all the interactions between two vertices from $G_i$ and $G_j$ need to go through the vertex $v$. The removing of $v$ would lead to the total disconnection between $G_i$ and $G_j$.

## III. SYSTEM CONFIGURATION

The simulation system consists of three networks: transmission network (*T*), communication network (*C*), and distribution network (*D*). *T* network is a synthetic Illinois regional 200-bus transmission system [22], which consists of 111 substations (nodes) that are named from node 1 to node 111. Twenty substations with the highest load are connected to distribution feeders. Two simplified feeder models from the 24 prototypical radial distribution feeder models are considered in this work, namely R5-12.47-1 and R5-12.47-2 [23]. The numbers of nodes are 142 and 67 for these two feeders, respectively. Each of the twenty substations is connected to an appropriate feeder model according to the size of its load. Among the 20 substations that have feeders attached, 17 of them are connected to R5-12.47-1 feeder, and 3 of them are connected to R5-12.47-2 feeder (Table I). Feeder nodes are named using two numbers separated by a dot as "*XYZ.ABC*". The number before the dot ("*XYZ*") represents the substation name, while the number after the dot ("*ABC*") represents the node index within the feeder. For example, node 65.046 shows that this feeder node is within a feeder attached to the substation 65 and the node index inside the feeder is 046. For each feeder, we assume 3 nodes are equipped with DERs, they are randomly selected among nodes other than the substation. Substations and DERs are assumed to have communication capabilities. Therefore, each of them has a corresponding node in the *C* network. Substation nodes in the *C* network share the same name with their corresponding names in the *T* network, and DER nodes in the *C* network share the same name with their corresponding names in the *D* network. In the *C* network, DERs are directly connected to their respective substations (Fig. 2 – 3), while the connections among substations are synthesized by randomly rewiring 10% of the *T* network edges. The resultant topology of *T* and *C* networks are shown in Fig. 1 with distribution-level nodes hidden for clarity.

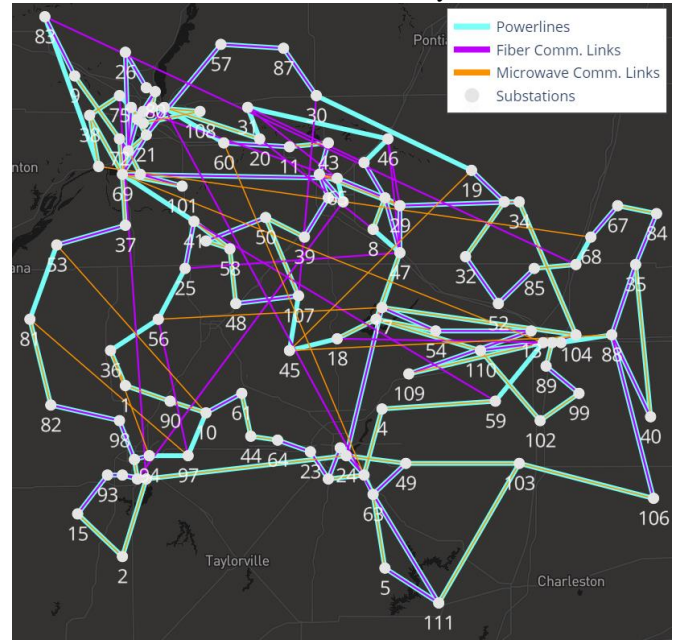

Fig. 1. Topologies of *T* and *C* Networks

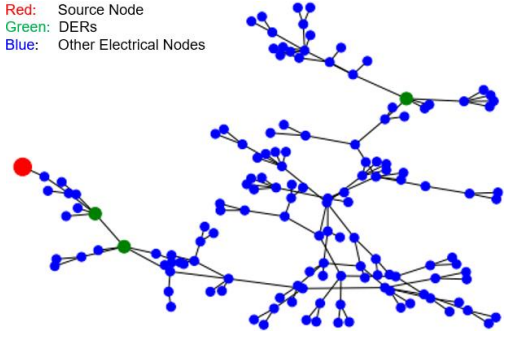

Fig. 2. Topology of D Network Feeder R5-12.47-1

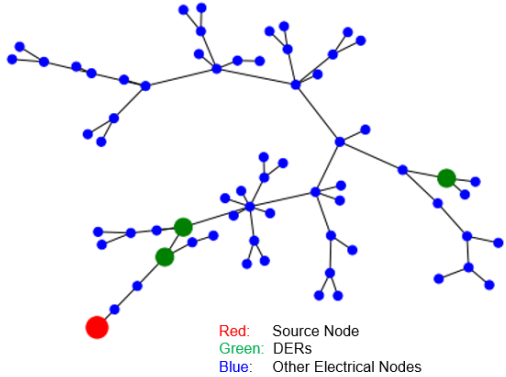

Fig. 3. Topology of *D* Network Feeder R5-12.47-2

TABLE I
SUBSTATION NODES THAT HAVE ATTACHED FEEDERS

| Node | Attached Feeder |
|------|-----------------|
| 65 | R5_12.47-1 |
| 70 | R5_12.47-1 |
| 18 | R5_12.47-1 |
| 92 | R5-12.47-2 |
| 96 | R5_12.47-1 |
| 12 | R5_12.47-1 |
| 6 | R5_12.47-1 |
| 7 | R5_12.47-1 |
| 24 | R5_12.47-1 |
| 78 | R5_12.47-1 |
| 105 | R5_12.47-1 |
| 14 | R5_12.47-1 |
| 91 | R5_12.47-1 |
| 73 | R5_12.47-1 |
| 27 | R5_12.47-1 |
| 108 | R5_12.47-1 |
| 48 | R5-12.47-2 |
| 75 | R5_12.47-1 |
| 13 | R5_12.47-1 |
| 60 | R5-12.47-2 |

## IV. GENERAL PROPERTIES OF *T&D&C* SYSTEM

The degree of a node, which is defined as the number of edges that are connected to the node, has been shown to be a good indicator of the network topological importance, especially its probability distribution ([5], [18]). For *T*, *D*, and *C* networks, majority of the nodes only have 1 or 2 edges (57.66%, 69.95%, and 65.51%, respectively), and only few nodes have higher number of degrees (> 5): 2.7%, 11.5%, and

2.1% for *T*, *C*, and *D* networks, respectively (Fig. 4), indicating that all 3 networks are vulnerable to targeted attacks on nodes with high degrees (hubs) and robust to random failures of nodes with low degrees. Among all three networks, *T* network has the highest average degree of 2.81 and *D* network has the lowest average degree of 1.98 (Table II). The highest degree is 8 for both *T* and *D* networks, and 9 for the *C* network.

The network efficiency, which is inversely proportional to the shortest distance in a network, indicates how efficiently energy and/or information can be exchanged within the network. The efficiency of a network, *G*, is defined as $E(G) = \frac{1}{N(N-1)}\sum_{i \neq j \in G}\frac{1}{d_{ij}}$, where $d_{ij}$ is the shortest path between node $i$ and $j$. The efficiency for the *T* network is slightly higher than the C network, 0.319 and 0.257, respectively, suggesting that the energy flow is more efficient within the *T* network than the information exchange within the *C* network. Because those 20 individual feeder networks in *D* network are not connected with each other, each node in *D* network only has shortest paths with nodes in its feeder network, while not with nodes in other feeder networks. As a result, the overall efficiency of the *D* network becomes very small and is only 0.012.

TABLE II
GENERAL PROPERTIES OF *T*, *D*, and *C* NETWORKS

| Network | N | L | <k> | k_max | E(G) |
|---------|---|---|-----|-------|------|
| T | 111 | 156 | 2.81 | 8 | 0.319 |
| C | 111(T)+72(DERs) | 226 | 2.47 | 9 | 0.257 |
| D | 2615 | 2595 | 1.98 | 8 | 0.012 |

*N* is the number of nodes; *L* is the number of edges; *<k>* is the average degree of a node; *k_max* is the degree of the node with the maximum number of connections (hub); *E(G)* is the network's efficiency.

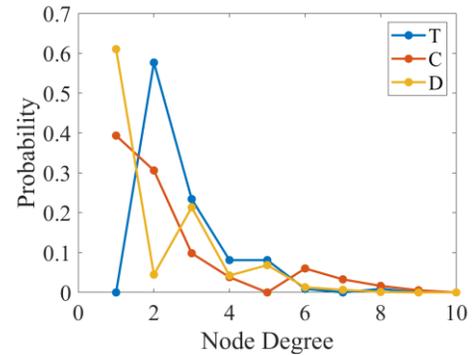

Fig. 4. Node Degree Probability Mass for *T*, *D*, and *C* Networks

## V. COMPARISON OF NETWORK ANALYSIS AND PHYSICS-BASED SIMULATION RESULTS

We first analyzed the *C* network and identified the top 10 critical nodes using the proposed network metrics, which are compared with the results from the ns-3 simulator [18] (Table III). For the ns-3 simulation, we sent packages between every two nodes in the *C* network and recorded package delays (*ΔT*) for all packages. The simulation was repeated 100 times, and the median package delays of the 100 simulations was calculated and recorded. Then we did the same simulation 100 times while removing each individual node in the network to assess how each node affected the package delays in the network.

TABLE III
Top 10 Critical Nodes based on Network Analysis and Ns-3 Simulation for the $C$ Network

| Rank | $c_v^j$ | $b_v^j$ | $\Delta E_G(v)$ | $\Delta T$ |
|------|---------|---------|-----------------|------------|
| 1 | 69 | 69 | 69 | 66 |
| 2 | 65 | 66 | 66 | 6 |
| 3 | 66 | 65 | 65 | 7 |
| 4 | 14 | 14 | 14 | 96 |
| 5 | 28 | 47 | 17 | 92 |
| 6 | 29 | 3 | 38 | 65 |
| 7 | 104 | 27 | 27 | 95 |
| 8 | 27 | 17 | 47 | 34 |
| 9 | 3 | 34 | 37 | 16 |
| 10 | 38 | 104 | 34 | 39 |

TABLE IV
Top 10 Critical Nodes based on Network Analysis and PowerModelsRestoration.jl Simulations for the $T$ Network

| Rank | | | | Cost of System Re-balancing |
|------|---------|---------|-----------------|------------|
| | $c_v^j$ | $b_v^j$ | $\Delta E_G(v)$ | |
| 1 | 29 | 69 | 69 | 95 |
| 2 | 28 | 65 | 65 | 3 |
| 3 | 65 | 16 | 17 | 69 |
| 4 | 47 | 17 | 24 | 16 |
| 5 | 16 | 24 | 47 | 17 |
| 6 | 17 | 47 | 95 | 84 |
| 7 | 66 | 29 | 34 | 35 |
| 8 | 34 | 28 | 16 | 62 |
| 9 | 69 | 34 | 3 | 65 |
| 10 | 3 | 104 | 29 | 104 |

Nodes were ranked based on how they impact the package delays using the median package delays from high to low.

The difference of how the shortest path distance was used led to the difference between the results from the network analysis and the ns-3 simulations. For example, although both the node efficiency drop in the network analysis and the package delays in ns-3 simulations used the shortest path distance between nodes as the input, the node efficiency drop in (3) was calculated using the reciprocal of the shortest path distance, while the package delays in ns-3 simulations were proportional to the length of the shortest path distance. To be more specific, in the network analysis results (Table III), node 69 is the top node using all three network metrics, which was expected by observing Fig. 1 where node 69 is in upper left region. The highest closeness centrality of node 69 indicates that it has the shortest path length on average to other nodes in the network, and the highest betweenness centrality of node 69 suggests that it falls on the shortest path between other pairs of nodes in the network. As a result, the removal of node 69 led to the highest efficiency drop of the network.

On the other hand, package delay is a unique characteristic between two communication nodes, which may separate it from other graph metrics; in comparison, electric potential is a unique characteristic of nodes in grid network. It should be noted that the highest ranked node was node 66 from the ns-3 simulations: the removal of node 66 caused the highest package delays than any other nodes; potential reasons includes its location in this network, as well as the simulation parameter settings/configurations in ns-3 simulations. For top ranked node from the ns-3 simulations, both nodes 65 and 66 were also top ranked nodes using three network metrics, but not for other top ranked nodes from the ns-3 simulations.

The maximum loading capability analysis for the $T$ network was done using *PowerModelsRestoration.jl*. For each node in the $T$ network, we simulated the power flow after single node removal, as well as removing the generators/loads attached to the node, if applicable. Then we calculated the ratio of system generation *vs* system load, which indicated the extra cost of the power system re-balancing. The top 10 ranked nodes from the *PowerModelsRestoration.jl* simulation were compared with the network analysis results for the $T$ network (Table IV). Their results agree quite well: 6 of the top ranked nodes from the *PowerModelsRestoration.jl* simulation are also among the top ranked nodes from the network node efficiency drop analysis.

This agreement verifies the applicability of network analysis results for planning or prioritizing the restoration procedure: giving higher priorities to top ranked nodes identified from the network efficiency can not only increase the overall network efficiency, but also improve the cost of re-balancing the system.

## VI. Interdependency Between $T\&D\&C$ Networks

### A. Interdependency Between $T$ and $C$ Networks

Fig. 5 shows histogram of the cross-efficiency drop when passing information from the $T$ network to the $C$ network, and from $C$ network to $T$ network, when removing each individual node. Clearly, removing most of nodes in both networks only cause less than 1% cross efficiency drop. Moreover, the removing of majority of nodes in both networks: 96.4% in $T$ network and 96.7% in $C$ network, only cause less than 3% efficiency drop. In $C$ network, the single node removal of 83% of nodes only cause efficiency drop of 1%. Table V lists the top 10 nodes in the $T$ network that can cause the highest cross-efficiency drop when being removed. The most critical nodes are nodes 69, 65, and 3, which would cause 6.21%, 5.97%, and 4.92% efficiency drop if being removed, respectively. In comparison, the removing of any other nodes would cause 3.61% or less efficiency drops. Among top 10 nodes, both nodes 69 and 3 are top ranked nodes for cross-closeness centrality and cross-betweenness centricity in the $T$ network, indicating that they are both the closest nodes and the most important information hub when passing information from $T$ network to $C$ network.

Because of the different topologies and edge weights between $C$ network and $T$ network, the identified critical nodes using cross-efficiency drop for $C$ network are different than the $T$ network: the top 5 nodes are 14, 27, 65, 69, and 75 (Table VI). Among all nodes in the $C$ network, removing each of the top 4 nodes on the list could cause over 4% cross-efficiency drop, which is higher than any other lower ranked nodes. There is also a strong overlapping of critical nodes identified using cross betweenness centrality and cross-efficiency drop. These two lists share 7 nodes between each other. In comparison, only 3 nodes identified using cross closeness centrality are on the top 10 node list identified using the cross-efficiency drop, which

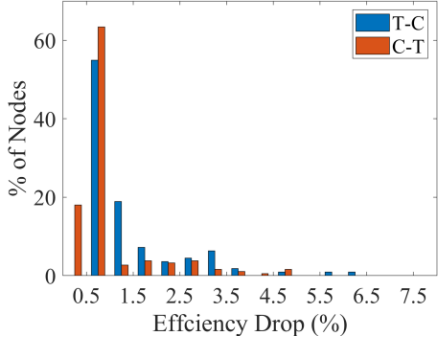

Fig. 5. Histogram of Cross-Efficiency Drop between $T$ and $C$ Networks

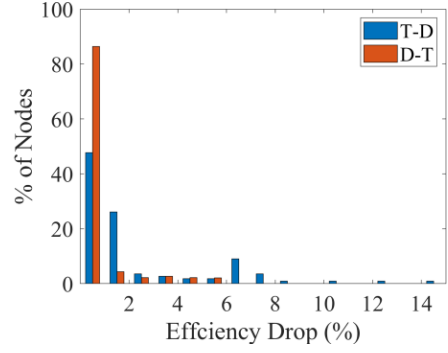

Fig. 6. Histogram of Cross-Efficiency Drop between $T$ and $D$ Networks

TABLE V

CRITICAL NODES IN THE $T$ NETWORK FOR THE INTERDEPENDENCY BETWEEN $T$&$C$ NETWORKS

| Rank | $T - C$ | | | | | |
|------|---------|---|---|---|---|---|
| | $c_v^I$ | | $b_v^I$ | | $\Delta E_G(v)$ | |
| | Node | Value | Node | Value | Node | Value |
| 1 | 69 | 0.444 | 69 | 2074 | 69 | 6.21% |
| 2 | 3 | 0.428 | 3 | 1134 | 65 | 5.97% |
| 3 | 65 | 0.422 | 16 | 1003 | 3 | 4.92% |
| 4 | 72 | 0.415 | 27 | 979 | 47 | 3.61% |
| 5 | 28 | 0.403 | 65 | 954 | 95 | 3.61% |
| 6 | 66 | 0.403 | 17 | 890 | 78 | 3.44% |
| 7 | 29 | 0.400 | 63 | 866 | 16 | 3.30% |
| 8 | 27 | 0.399 | 91 | 823 | 27 | 3.28% |
| 9 | 78 | 0.399 | 95 | 821 | 29 | 3.27% |
| 10 | 47 | 0.388 | 92 | 790 | 66 | 3.21% |

TABLE VII

CRITICAL NODES IN THE $T$ NETWORK FOR THE INTERDEPENDENCY BETWEEN $T$&$D$ NETWORKS

| Rank | $T - D$ | | | | | |
|------|---------|---|---|---|---|---|
| | $c_v^I$ | | $b_v^I$ | | $\Delta E_G(v)$ | |
| | Node | Value | Node | Value | Node | Value |
| 1 | 65 | 0.073 | 65 | 6531 | 65 | 14.08% |
| 2 | 28 | 0.073 | 69 | 4891 | 24 | 12.63% |
| 3 | 29 | 0.073 | 24 | 4720 | 69 | 10.42% |
| 4 | 47 | 0.073 | 16 | 3577 | 17 | 8.65% |
| 5 | 16 | 0.073 | 17 | 3355 | 91 | 7.60% |
| 6 | 66 | 0.073 | 3 | 3328 | 27 | 7.33% |
| 7 | 17 | 0.072 | 47 | 2889 | 105 | 7.13% |
| 8 | 69 | 0.072 | 12 | 2729 | 78 | 7.01% |
| 9 | 3 | 0.071 | 91 | 2654 | 6 | 6.99% |
| 10 | 72 | 0.071 | 27 | 2496 | 12 | 6.94% |

TABLE VI

CRITICAL NODES IN THE $C$ NETWORK FOR THE INTERDEPENDENCY BETWEEN $T$&$C$ NETWORKS

| Rank | $C - T$ | | | | | |
|------|---------|---|---|---|---|---|
| | $c_v^I$ | | $b_v^I$ | | $\Delta E_G(v)$ | |
| | Node | Value | Node | Value | Node | Value |
| 1 | 69 | 0.403 | 14 | 14091 | 14 | 4.90% |
| 2 | 3 | 0.388 | 27 | 13135 | 27 | 4.84% |
| 3 | 65 | 0.378 | 65 | 9509 | 65 | 4.63% |
| 4 | 65.046 | 0.378 | 75 | 7825 | 69 | 4.20% |
| 5 | 65.045 | 0.378 | 7 | 7103 | 75 | 3.86% |
| 6 | 65.033 | 0.378 | 91 | 5195 | 7 | 3.64% |
| 7 | 72 | 0.378 | 78 | 4757 | 78 | 3.15% |
| 8 | 28 | 0.368 | 18 | 4650 | 96 | 3.06% |
| 9 | 29 | 0.366 | 24 | 3627 | 91 | 3.04% |
| 10 | 27 | 0.365 | 60 | 3481 | 6 | 2.84% |

TABLE VIII

CRITICAL NODES IN THE $D$ NETWORK FOR THE INTERDEPENDENCY BETWEEN $T$&$D$ NETWORKS

| Rank | $D - T$ | | | | | |
|------|---------|---|---|---|---|---|
| | $c_v^I$ | | $b_v^I$ | | $\Delta E_G(v)$ | |
| | Node | Value | Node | Value | Node | Value |
| 1 | 65.042 | 0.282 | 78.032 | 4705 | 65.042 | 6.29% |
| 2 | 65.001 | 0.282 | 24.032 | 4705 | 65.001 | 6.05% |
| 3 | 65.041 | 0.282 | 70.032 | 4705 | 27.042 | 5.89% |
| 4 | 24.042 | 0.251 | 105.032 | 4705 | 7.042 | 5.83% |
| 5 | 24.001 | 0.251 | 96.032 | 4705 | 65.041 | 5.80% |
| 6 | 24.041 | 0.251 | 18.032 | 4705 | 24.042 | 5.79% |
| 7 | 7.042 | 0.244 | 14.032 | 4705 | 6.042 | 5.78% |
| 8 | 7.001 | 0.244 | 7.032 | 4705 | 27.001 | 5.70% |
| 9 | 7.041 | 0.244 | 27.032 | 4705 | 78.042 | 5.67% |
| 10 | 6.042 | 0.239 | 65.032 | 4705 | 7.001 | 5.65% |

suggests that node with higher cross betweenness centrality (the information hub) is more critical than nodes with higher cross closeness centrality (closer) when passing information from $C$ network to $T$ network. However, using cross betweenness centrality alone may not find all critical nodes. For example, although node 69 is not on the top 10 nodes list using the cross betweenness centrality, it is still one of the critical nodes that can cause higher cross-efficiency drop when being removed.

### B. Interdependency Between $T$ and $D$ Networks

Fig. 6 shows the histogram of cross-efficiency drop when passing information from $T$ network to $D$ network, and from $D$ network to $T$ network, when removing each individual node. Because of the tree topology of the $D$ network, the removing of a single node in the $D$ network would only cause less than 1% cross-efficiency drop for 86.73% of nodes. The removing of any of the top 10 nodes in the $T$ network could result in the cross-efficiency drop ranging from 6.94% to 14.08% (Table passing informing from *C* network to *D* network, and from *D* network to *C* network, when removing each individual node. The removing of any of the top 10 nodes in *C* network could cause efficiency drop ranging from 8.05% to 10.83% (Table IX). The comparison of nodes with/without feeders in the *C* network shows that the median cross-efficiency drops for removing nodes with feeders are much higher than removing nodes without feeders in *C* network: 7.86% and 0.82%, respectively (Fig. 9). The removing of any nodes with feeders would cause the cross-efficiency drop > 4%, except for the node 48. For nodes without feeders, although the removing of most of them would only cause the cross-efficiency drop < 4%, the removing of node 69 would cause the cross-efficiency drop 7.5%.

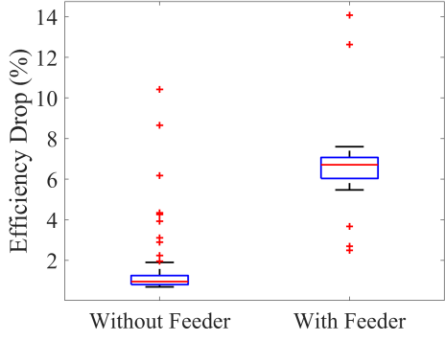

Fig. 7. Boxplot of Cross-Efficiency Drop from the *T* Network to the *D* Network for Nodes in the *T* network without and with Feeders

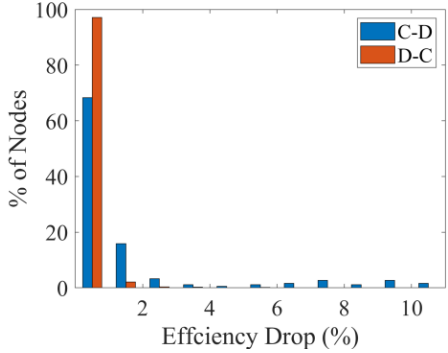

Fig. 8. Histogram of Efficiency Drop between *C* and *D* Networks

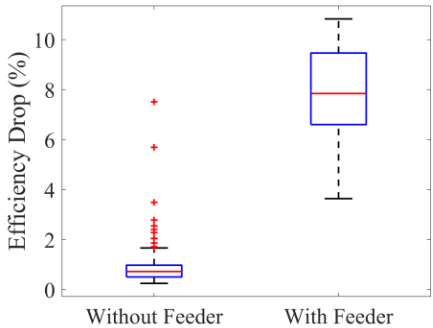

Fig. 9. Boxplot of Cross-Efficiency Drop from the *C* Network to the *D* Network for Nodes in the *C* network with and without Feeders

TABLE IX
CRITICAL NODES IN THE *C* NETWORK FOR THE INTERDEPENDENCY BETWEEN
*C&D* NETWORKS

| Rank | $c_v^j$ | | $b_v^j$ | | $\Delta E_G(v)$ | |
|---|---|---|---|---|---|---|
| | Node | Value | Node | Value | Node | Value |
| 1 | 69 | 0.128 | 7 | 78759 | 14 | 10.83% |
| 2 | 65 | 0.126 | 27 | 73698 | 7 | 10.78% |
| 3 | 65.046 | 0.126 | 14 | 73638 | 65 | 10.57% |
| 4 | 65.045 | 0.126 | 18 | 67766 | 6 | 9.71% |
| 5 | 65.033 | 0.126 | 6 | 64923 | 75 | 9.64% |
| 6 | 14 | 0.123 | 13 | 55752 | 27 | 9.29% |
| 7 | 14.003 | 0.123 | 12 | 52047 | 91 | 9.16% |
| 8 | 14.029 | 0.123 | 108 | 48001 | 12 | 9.15% |
| 9 | 14.043 | 0.123 | 96 | 33484 | 96 | 8.37% |
| 10 | 14.02 | 0.123 | 75 | 32598 | 18 | 8.05% |

TABLE X
CRITICAL NODES IN THE *D* NETWORK FOR THE INTERDEPENDENCY BETWEEN
*C&D* NETWORKS

| Rank | $c_v^j$ | | $b_v^j$ | | $\Delta E_G(v)$ | |
|---|---|---|---|---|---|---|
| | Node | Value | Node | Value | Node | Value |
| 1 | 65.045 | 0.274 | 91.032 | 7823 | 6.023 | 5.71% |
| 2 | 65.122 | 0.274 | 78.032 | 7686 | 75.023 | 5.31% |
| 3 | 65.123 | 0.274 | 24.02 | 6998 | 6.03 | 5.02% |
| 4 | 65.046 | 0.274 | 70.02 | 6998 | 75.03 | 4.97% |
| 5 | 65.033 | 0.274 | 105.02 | 6998 | 7.03 | 3.90% |
| 6 | 65.131 | 0.274 | 65.02 | 6998 | 7.023 | 3.77% |
| 7 | 65.132 | 0.274 | 73.02 | 6998 | 12.023 | 3.75% |
| 8 | 65.087 | 0.274 | 60.01 | 6977 | 12.03 | 3.73% |
| 9 | 65.088 | 0.274 | 48.01 | 6859 | 105.046 | 3.25% |
| 10 | 65.047 | 0.272 | 6.023 | 6797 | 105.047 | 3.05% |

Like the interdependency between *T* and *D*, the removing 97.40% of *D* nodes only cause less than 1% cross-efficiency drop, and the max cross-efficiency drop is 5.6% (Table X), indicating that the removing of an individual node in *D* network doesn't affect much of the overall cross-efficiency, which is expected because each feeder network has 3 *DERs*. When one *DER* node is removed, information can still be passed from *D* network to *C* network through the other two *DER* nodes. Unlike identified top nodes from *D* network to *T* network, which are

VII). In comparison, the max cross-efficiency drop is less than 6.5% in the *D* network and all identified top nodes are either the feeder head (65.024, 27.042, ...) or the node directly connected with the feeder head (65.001, 27.001, ...) in the *D* network (Table VIII). The removing of those top three nodes in the *T* network: 65, 24, and 69, could lead to over 10% cross-efficiency drop from the *T* network to the *D* network. The median efficiency drops for removing nodes with feeders are much higher than removing nodes without feeders in *T* network: 6.71% and 0.95%, respectively (Fig. 7). However, two of top 10 nodes: node 69 and 17 have no loads/feeders and removing either of them could cause higher cross-efficiency drop than the remaining 18 nodes that have feeders.

### C. Interdependency Between *C* and *D* Networks

Fig. 8 shows the histogram of cross-efficiency drop of all root nodes or nodes next to root nodes (Table VIII), identified top nodes in *D* network using the cross efficiency drop include both *DER*s (6.023, 6.03, …) and *non-DER*s (105.047, …). Another notable observation is that because of the tree-like nature of the feeder graph, removing some critical *non-DER* nodes could cut the graph into disconnected pieces and significantly increase the length of shortest paths from *D* to *C* for some nodes, which in turn increases the overall cross-efficiency drop.

| Rank | Edge Weight = 1 | | | Physical-based Edge Weight | | |
|---|---|---|---|---|---|---|
| | $\Delta E_G(v)$ (Node) | $\Delta E_G(v)$ (%) | Voltage (kV) | $\Delta E_G(v)$ (Node) | $\Delta E_G(v)$ (%) | Voltage (kV) |
| 1 | 69 | 13.62% | 230 | 69 | 15.37% | 230 |
| 2 | 65 | 5.86% | 230 | 65 | 10.56% | 230 |
| 3 | 24 | 5.19% | 230 | 17 | 9.00% | 230 |
| 4 | 17 | 4.41% | 230 | 24 | 7.98% | 230 |
| 5 | 41 | 4.22% | 115 | 47 | 5.67% | 230 |
| 6 | 34 | 4.00% | 230 | 95 | 5.49% | 230 |
| 7 | 95 | 3.57% | 230 | 34 | 5.22% | 230 |
| 8 | 33 | 3.07% | 115 | 16 | 4.97% | 230 |
| 9 | 29 | 2.99% | 230 | 3 | 4.78% | 230 |
| 10 | 3 | 2.93% | 230 | 29 | 4.28% | 230 |

| Rank | Edge Weight = 1 | | | Physical-based Edge Weight | | |
|---|---|---|---|---|---|---|
| | $\Delta E_G(v)$ (Node) | $\Delta E_G(v)$ (%) | Voltage (kV) | $\Delta E_G(v)$ (Node) | $\Delta E_G(v)$ (%) | Voltage (kV) |
| 1 | 69 | 2.74% | 230 | 69 | 6.21% | 230 |
| 2 | 65 | 2.04% | 230 | 65 | 5.97% | 230 |
| 3 | 27 | 1.95% | 230 | 3 | 4.92% | 230 |
| 4 | 41 | 1.74% | 115 | 47 | 3.61% | 230 |
| 5 | 78 | 1.73% | 115 | 95 | 3.61% | 230 |
| 6 | 105 | 1.60% | 115 | 78 | 3.44% | 115 |
| 7 | 75 | 1.59% | 115 | 16 | 3.30% | 230 |
| 8 | 7 | 1.58% | 115 | 27 | 3.28% | 230 |
| 9 | 14 | 1.57% | 115 | 29 | 3.27% | 230 |
| 10 | 60 | 1.51% | 115 | 66 | 3.21% | 230 |

| Rank | *T* Network Only | | *T&C* | |
|---|---|---|---|---|
| | $\Delta E_G(v)$ (Node) | Voltage (kV) | $\Delta E_G(v)$ (Node) | Voltage (kV) |
| 1 | 69 | 230 | 69 | 230 |
| 2 | 65 | 230 | 65 | 230 |
| 3 | 17 | 230 | 3 | 230 |
| 4 | 24 | 230 | 47 | 230 |
| 5 | 47 | 230 | 95 | 230 |
| 6 | 95 | 230 | 78 | 115 |
| 7 | 34 | 230 | 16 | 230 |
| 8 | 16 | 230 | 27 | 230 |
| 9 | 3 | 230 | 29 | 230 |
| 10 | 29 | 230 | 66 | 230 |

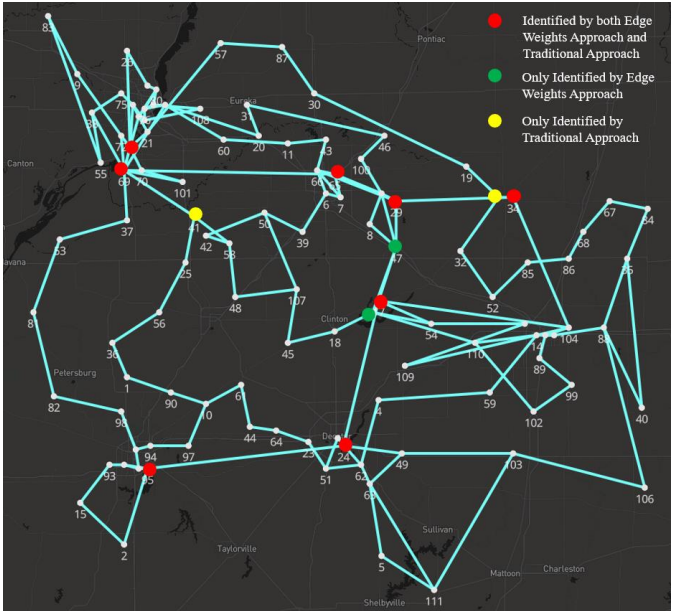

Fig. 10. Locations of Identified Top 10 Nodes using Both the Proposed Physical-based Edge Weight approach and the Traditional Approach.

## VII. COMPARISON OF THE PROPOSED APPROACH AND EXISTING APPROACHES

To illustrate the difference between the proposed physical-based weight approach and the traditional approach that set all edge weights to 1 [17], we first compared the identified top 10 nodes from the node efficiency drop results using both approached for the single *T* network. As shown in Table XI, while results using both approaches were comparable: 8 identified nodes were shared by both approaches and only two nodes were different from between them (see Fig. 10). However, the proposed edge weight approach clearly gave more priorities to nodes with higher voltages: all identified 10 critical nodes have 230 kV nominal voltage, while the two nodes that were only identified using the traditional approach have lower nominal voltage (115 kV). The calculated efficiency drops were also higher using the proposed approach than the traditional approach, indicating that the identified nodes were more important than other lower ranked nodes when using the proposed edge weight approach. The difference was also observed in identified critical nodes for the interdependency from the *T* network to the *C* network (Table XII). Although nodes 69 and 65 were identified as the top two nodes using both approaches, 7 out of 10 critical nodes identified using the traditional approach have 115 kV nominal voltage. Using the proposed edge weight approach, only 1 critical node (78) had nominal voltage 115 kV and all other 9 nodes had higher nominal voltage 230 kV. Similarly, the calculated efficiency drops were also higher for critical nodes identified using the proposed edge weight approach than using the traditional approach.

Table XIII shows the critical nodes in the single *T* network and T&*C* network. Although both node sets share 7 same nodes and have the same top 2 nodes, ranks of the rest node are different, which suggests that the importance of each individual node may be different in multilayer networks than in a single layer network. For example, node 17 is ranked at the third place in the single *T* network, but it is outside the top 10 critical nodes list for the interdependency from the *T* network to the *C* network. This indicates it is important to consider the network mutual interdependence and quantify the different node importance in interdependent *T&D&C* networks.

## VIII. Conclusions

In this paper, we proposed a physics-embedded multilayer network modeling framework and several cross-network interdependency metrics for analyzing the component important and their interdependence within an interdependent cyber-physical system that comprises three different types of networks: a power transmission network (*T*), multiple distribution networks (*D*) and a communication network (*C*). Our physical-based approach for the network edge weight calculation incorporated both line impedances and node voltages for both the *T&D* networks.

We also tried to answer an important question: to what extent topological network analysis results are consistent with physical-based analysis results? We perform two comparison studies. With this physics-embedding network modeling, the network analysis and the physical-model based maximum loading capacity analysis for the *T* network showed comparable results. This validates the effectiveness of network theory based analysis for electrical power systems. However, for the communication system, there are still some notable differences between NS-3 simulation and network theory results, in part due to inadequate modeling of communication system physical properties in the graph modeling.

The results also showed that the network topologies have direct impacts on the component importance, which is one of the most important factors for the system restoration. It is not surprising that nodes with an attached feeder are generally more critical than nodes without an attached feeder, especially for the interdependency from the *T* network or the *C* network to the *D* network. However, the multilayer network analysis also shows that some nodes that don't have attached feeders, such as nodes 69 and 17 in the *T* network and node 66 in the C network, are also critical components due to their unique locations in the system topology. For the tree-like networks, such as the *D* network, the results show that the root node or the *DER* nodes are more critical than other leaf nodes for the interdependency between the *D* network and the *T* network or the C network. Besides the root node or the *DER* nodes, some *non-DER* nodes can also play critical roles because removing them could significantly increase the length of shortest path from the *D* network to the *C* network. For the interdependence between the *T* network and the *C* network, although sharing the same set of nodes, the results showed that different set of critical nodes exist in each network because of the different topologies and underlying physics of these two networks.

Future research directions of interest include: 1) coordinated node importance analysis could be extended by incorporating the active and reactive power flow in both the *T* network and the *D* network [24], and corresponding directed graphs can be formulated and analyzed; 2) improved representation of communication system physical properties in the graph modeling; 3)besides the node importance, we are also planning to apply the proposed framework for identifying critical edges/transmission lines in the studied interdependent *T&D&C* system; 4) cascading failures ([25-27]) may be considered in the simulation to provide more comprehensive interdependency analysis for system-level high impact low frequency events.

## Acknowledgement

The Pacific Northwest National Laboratory is operated by Battelle for the U.S. Department of Energy (DOE) under Contract DE-AC05-76RL01830. This work was supported by DOE Office of Electricity, Advanced Grid Modeling (AGM) Program.

## References

[1] Ulusan A, Ergun O. Restoration of services in disrupted infrastructure systems: A network science approach. PloS one. 2018;13(2).

[2] Gengfeng LI, Huang G, Zhaohong BI, Yanling LI, Huang Y. Component importance assessment of power systems for improving resilience under wind storms. Journal of Modern Power Systems and Clean Energy. 2019 Jul 13;7(4):676-87.

[3] Abedi A, Gaudard L, Romerio F. Review of major approaches to analyze vulnerability in power system. Reliability Engineering & System Safety. 2019 Mar 1;183:153-72.

[4] Chu CC, Iu HH. Complex networks theory for modern smart grid applications: A survey. IEEE Journal on Emerging and Selected Topics in Circuits and Systems. 2017 Apr 24;7(2):177-91.

[5] Albert R, Albert I, Nakarado GL. Structural vulnerability of the North American power grid. Physical review E. 2004 Feb 26;69(2):025103.

[6] Kinney R, Crucitti P, Albert R, Latora V. Modeling cascading failures in the North American power grid. The European Physical Journal B-Condensed Matter and Complex Systems. 2005 Jul 1;46(1):101-7.

[7] Crucitti P, Latora V, Marchiori M. Model for cascading failures in complex networks. Physical Review E. 2004 Apr 29;69(4):045104.

[8] Bompard E, Wu D, Xue F. Structural vulnerability of power systems: A topological approach. Electric Power Systems Research. 2011 Jul 1;81(7):1334-40.

[9] Luo XS, Zhang B. Analysis of cascading failure in complex power networks under the load local preferential redistribution rule. Physica A: Statistical Mechanics and its Applications. 2012 Apr 15;391(8):2771-7.

[10] Wei Z, Liu J. Research on the electric power grid vulnerability under the directed-weighted topological model based on Complex Network Theory. In2010 International Conference on Mechanic Automation and Control Engineering 2010 Jun 26 (pp. 3927-3930). IEEE.

[11] Rinaldi SM, Peerenboom JP, Kelly TK. Identifying, understanding, and analyzing critical infrastructure interdependencies. IEEE control systems magazine. 2001 Dec;21(6):11-25.

[12] Pederson P, Dudenhoeffer D, Hartley S, Permann M. Critical infrastructure interdependency modeling: a survey of US and international research. Idaho National Laboratory. 2006 Aug 1;25:27.

[13] Buldyrev SV, Parshani R, Paul G, Stanley HE, Havlin S. Catastrophic cascade of failures in interdependent networks. Nature. 2010 Apr;464(7291):1025-8.

[14] Setréus J, Hilber P, Arnborg S, Taylor N. Identifying critical components for transmission system reliability. IEEE Transactions on Power Systems. 2012 Apr 3;27(4):2106-15.

[15] Donges JF, Schultz HC, Marwan N, Zou Y, Kurths J. Investigating the topology of interacting networks. The European Physical Journal B. 2011 Dec 1;84(4):635-51.

[16] Wang S, Hong L, Chen X. Vulnerability analysis of interdependent infrastructure systems: A methodological framework. Physica A: Statistical Mechanics and its applications. 2012 Jun 1;391(11):3323-35.

[17] Milanović JV, Zhu W. Modeling of interconnected critical infrastructure systems using complex network theory. IEEE Transactions on Smart Grid. 2017 Feb 7;9(5):4637-48.

[18] Fan X, Aksoy S, Wang D, Huang Q, Ogle J, Tbaileh A, Huang R. Automated Realistic Testbed Synthesis for Power System Communication Networks based on Graph Metrics. In2020 IEEE Power & Energy Society Innovative Smart Grid Technologies Conference (ISGT) 2020 Feb 17 (pp. 1-5). IEEE.

[19] Riley G.F., Henderson T.R. (2010) The *ns-3* Network Simulator. In: Wehrle K., Güneş M., Gross J. (eds) Modeling and Tools for Network Simulation. Springer, Berlin, Heidelberg

[20] Noah Rhodes, David Fobes, Carleton Coffrin, and Line Roald, PowerModelsRestoration.jl: An Open-Source Framework for Exploring Power Network Restoration Algorithms, arXiv:2004.13177

[21] C. Coffrin, R. Bent, B. Tasseff, K. Sundar and S. Backhaus, "Relaxations of AC Maximal Load Delivery for Severe Contingency Analysis," in *IEEE Transactions on Power Systems*, vol. 34, no. 2, pp. 1450-1458, March 2019, doi: 10.1109/TPWRS.2018.2876507.

[22] Birchfield, A.B., Xu, T., Gegner, K.M., Shetye, K.S. and Overbye, T.J., 2016. Grid structural characteristics as validation criteria for synthetic networks. *IEEE Transactions on power systems*, *32*(4), pp.3258-3265.

[23] Schneider, K.P., Chen, Y., Chassin, D.P., Pratt, R.G., Engel, D.W. and Thompson, S.E., 2008. Modern grid initiative distribution taxonomy final report (No. PNNL-18035). Pacific Northwest National Lab.(PNNL), Richland, WA (United States).

[24] Q. Huang and V. Vittal, "Integrated Transmission and Distribution System Power Flow and Dynamic Simulation Using Mixed Three-Sequence/Three-Phase Modeling," in *IEEE Transactions on Power Systems*, vol. 32, no. 5, pp. 3704-3714, Sept. 2017

[25] Korkali, M., Veneman, J.G., Tivnan, B.F., Bagrow, J.P. and Hines, P.D., 2017. Reducing cascading failure risk by increasing infrastructure network interdependence. *Scientific reports*, *7*, p.44499.

[26] Boccaletti, S., Bianconi, G., Criado, R., Del Genio, C.I., Gómez-Gardenes, J., Romance, M., Sendina-Nadal, I., Wang, Z. and Zanin, M., 2014. The structure and dynamics of multilayer networks. *Physics Reports*, *544*(1), pp.1-122.

[27] Fan, X., Agrawal, U., Davis, S., O'Brien, J., Etingov, P., Nguyen, T., Makarov, Y. and Samaan, N., 2020, August. Bulk Electric System Protection Model Demonstration with 2011 Southwest Blackout in DCAT. In *2020 IEEE Power & Energy Society General Meeting (PESGM)* (pp. 1-5). IEEE.